\documentstyle[aps,prl,psfig,special]{revtex}

\def\text#1{{\rm #1}}
\def\epsilon{\varepsilon}

\begin{document}
\author{Fabian Braun, and Jan von Delft} 
\title{\vspace*{0.7cm}
  Fixed-$N$ Superconductivity:\\
  The Exact Crossover from the Bulk to the Few-Electron Limit}
\address{Institut f\"ur Theoretische Festk\"orperphysik, Universit\"at
  Karlsruhe, D--76128 Karlsruhe, Germany} 
\maketitle

\begin{abstract}
  We use two truly canonical approaches to describe superconductivity in
  ultrasmall metallic grains: (a) a variational fixed-$N$ projected
  BCS-like theory and (b) the exact solution of the model Hamiltonian
  developed by Richardson in context with Nuclear Physics.  Thereby we
  obtain a description of the entire crossover from the bulk BCS regime
  (mean level spacing $d\ll$ bulk gap $\tilde\Delta$) to the
  ``fluctuation-dominated'' few-electron regime ($d\gg\tilde\Delta$). A
  wave-function analysis shows in detail how the BCS limit is recovered and
  how for $d\gg\tilde\Delta$ pairing correlations become delocalized in
  energy space.
\end{abstract}
\vspace*{0.4cm}
\narrowtext
\section{Introduction}
When a system of (correlated) electrons is sufficiently small, the
electronic spectrum becomes discrete.  Such a discrete spectrum was
directly measured for the first time by Ralph, Black and Tinkham (RBT)
\cite{Ralph-95,Ralph-97}, for ultrasmall Al grains. This allowed them to
study the nature of pairing correlations in a small superconductor in
unprecedented detail.  These experiments gave new actuality to an old and
fundamental question: What is the lower size limit for superconductivity?
Besides Anderson's prediction \cite{Anderson-59} that
superconductivity breaks down once the single-particle mean level
spacing $d$ becomes larger than the (bulk) superconducting gap
$\tilde\Delta$, the finite size of a superconducting grain also
manifests itself in its large charging energy, which  effectively suppresses
electron number fluctuation on the grain. Hence superconductivity on
small grains has to be formulated in a manifestly canonical way.

After briefly introducing the experiments and a toy model which captures
their essential features, we show how the entire crossover between the bulk
BCS-like regime and the few-electron regime can be described theoretically
by fixing the electron number on the grain and using either a projected BCS
approach or, even better, a long-forgotten {\em exact solution of the
  reduced BCS model Hamiltonian}. Both the projected BCS approach and the
exact solution enable us (i) to significantly improve previous g.c.\ upper
bounds on ground state energies,
\cite{Golubev-94,vonDelft-96,Braun-97,Braun-99}, in the latter case by
giving the exact result; (ii) to find in the crossover regime a remnant of
the ``break-down of superconductivity'' obtained in g.c.\ studies, at which
the condensation energy changes from being extensive to practically
intensive; and (iii) to study this change by an {\em explicit wave-function
  analysis}, which shows in detail how the BCS limit is recovered for $d\ll
\tilde \Delta$, and how for $d \gg \tilde \Delta$ pairing correlations
delocalize in energy space.

\subsection{Spectroscopic Gap in an Ultrasmall Superconducting Grain}

In RBT's experiments, an ultrasmall grain was used as central island in a
single-electron transistor: it was connected via tunnel barriers to
external leads and capacitively coupled to a gate, and its electronic
spectrum determined by measuring the tunnel current through the grain as a
function of transport voltage at a fixed temperature of 50mK. For a typical
grain the radius was $r\approx 5$nm, and the crude order-of-magnitude
free-electron estimate $d= 2 \pi^2 \hbar^2 / (m k_F \mbox{Vol})$ for the
mean level spacing near $\epsilon_F$ yields $d \simeq 0.5$meV. The grain's
charging energy was about $E_C = e^2/2C_{\rm total} = 50$meV and therefore
much larger than all other energy scales, such as the Aluminum bulk gap
($\tilde\Delta \simeq 0.4$meV), typical values of the transport voltage ($V
\,{\stackrel{\scriptscriptstyle <}{\scriptscriptstyle \sim}}\, 2$mV) and
the temperature.

\begin{figure}[p]
  \begin{center}
    \leavevmode
    \psfig{figure=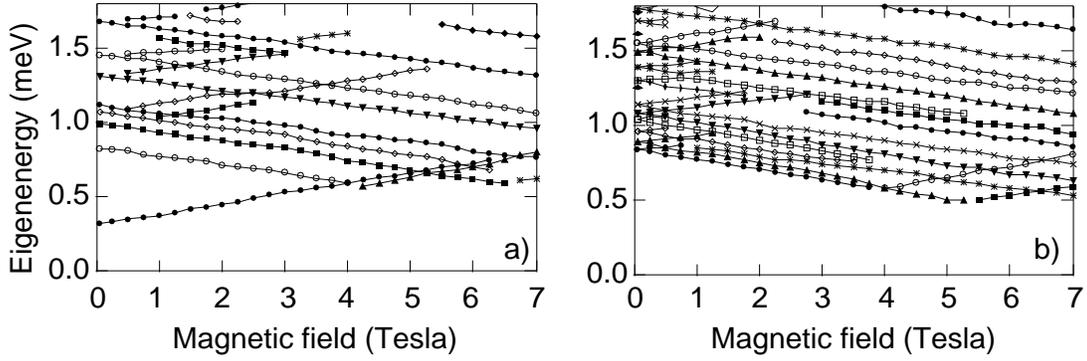,width=0.8\hsize}\vspace*{1ex}
    \caption{RBT's experimental tunneling spectra 
      \protect\cite{Ralph-97}.  The distances between
      lines give the fixed-$N$ excitation spectra of the same
      grain containing (a) an even and (b) an
      odd number of electrons, as function of magnetic field.}
    \label{fig:experimental-spectra}
  \end{center}
\end{figure}

The most remarkable feature of the experimental tunneling spectrum, shown
in Fig.~\ref{fig:experimental-spectra}, is the presence of a clear
spectroscopic gap for the grain with even electron number and its absence
for the odd grain.  This reveals the presence of pairing correlations: in
even grains, all excited states involve at least two BCS quasi-particles
and hence lie significantly above the ground state, whereas odd grains {\em
  always\/} have at least one quasi-particle and excitations need not
overcome an extra gap.  

The charging energy, being the largest energy scale of the system, strongly
suppresses particle number fluctuations on the grain  and hence the discrete
energies measured in RBT's experiments essentially correspond to the
eigenspectrum of a grain with {\em fixed electron number $N$}.  We therefore
consider below an ultrasmall grain {\em completely isolated\/} from the rest
of the world, e.g.\ by infinitely thick oxide barriers. Our main goal will be
to elucidate, within a canonical framework, the nature of the pairing
correlations in an ultrasmall grain at $T=0$.

\section{Modelling and First Approximation}
To investigate the influence of pairing correlations, on the excitation
spectrum of an ultrasmall grain, we model the grain by a reduced
BCS-Ha\-mil\-ton\-ian. It has been used before to describe small
superconducting grains \cite{vonDelft-96,Braun-97,Braun-99,Matveev-97} and
was phenomenologically successful for $d\le \tilde \Delta$
\cite{Braun-97,Braun-99}, but probably is unrealistically simple for $d \gg
\tilde \Delta$, for which it should rather be viewed as toy model:
\begin{equation}
  \label{eq:hamiltonian}
  H = \sum_{j=0,\sigma}^{N-1}\epsilon_j c^\dagger_{j\sigma}c_{j\sigma} 
  - \lambda \, d \sum_{j,j'=0}^{N-1} c^\dagger_{j+}c^\dagger_{j-}c_{j'-} c_{j'+}.
\end{equation}
The $c^\dagger_{j \pm}$ create electrons in free time-reversed
single-particle-in-a-box states $|j,\pm\rangle$, with discrete, uniformly
spaced, doubly degenerate eigenenergies $\epsilon_j= j d +\epsilon_0$.
The interaction scatters only time-reversed pairs of electrons within
$\omega_D$ of $\epsilon_F$.  Its dimensionless strength $\lambda$ is
related to the two material parameters $\tilde\Delta$ and $\omega_D$ via the
bulk gap equation $\sinh(1/\lambda) = \omega_D/\tilde\Delta$. We chose
$\lambda=0.224$, close to that of Al \cite{Braun-99}.  The level spacing $d$
determines the number $N=2\omega_D/d$ of levels, taken symmetrically around
$\epsilon_F$, within the cutoff; electrons outside the cutoff remain
unaffected by the interaction and are thus neglected throughout. 

\subsection{Grand-Canonical BCS Approach}
The most direct and easiest theoretical approach \cite{vonDelft-96} to
describing an ultrasmall grain simply uses the well-known
grand-canonical (g.c.) variational BCS ansatz for the ground state of an even
or odd grain (subscript $p = 0$ or 1, respectively):
\begin{eqnarray}
  \label{eq:BCSground}
  \begin{array}{rcccl}
    |\text{BCS}\rangle_0 & = & & \displaystyle
    \prod_j &\!\! (u_j + v_j c^\dagger_{j+}c^\dagger_{j-})\,|\text{Vac}\rangle\\
    |\text{BCS}\rangle_1 & = & c^\dagger_{j_{\text{odd}}} & \displaystyle
    \prod_{j\neq j_{\text{odd}}} & \!\!
    (u_j + v_j c^\dagger_{j+}c^\dagger_{j-})\,|\text{Vac}\rangle
  \end{array} \\
  \quad (u^2_j + v^2_j = 1) \; .\nonumber 
\end{eqnarray}          
$v_j$ and $u_j$ are the amplitudes that level $j$ is doubly
occupied or empty, respectively.  Note that on an odd grain one electron
necessarily is unpaired; to minimize its kinetic energy, it is put at the
Fermi energy ($\varepsilon_{j_{odd}} = \varepsilon_F$).  Minimizing the
energy expectation value ${\cal E}_p^{\text{GC}}=\langle\text{BCS}|\hat
H|\text{BCS}\rangle$ ($p=0,1$) with respect to $u_j$ and $v_j$ yields the
even and odd ``gap equations'' (at $T=0$):
\begin{eqnarray}
  \label{eq:gap-eq-even}
  \frac1\lambda = d \sum_j \frac1{\sqrt{\epsilon_j^2+\Delta_0^2(d)}}, \,
  \frac1\lambda = d \sum_{j\neq j_{\text{odd}}}
  \frac1{\sqrt{\epsilon_j^2+\Delta_1^2(d)}}. 
\end{eqnarray}
These are solved for the even and odd pairing parameters $\Delta_{0}$ and
$\Delta_{1}$ as a function of level-spacing $d$. Note that $j_{\text{odd}}$
is excluded from the odd sum.

As predicted in 1959 by Anderson, \cite{Anderson-59} it turned out
\cite{vonDelft-96} that above a critical level spacing the gap equation
ceases to have a non-trivial solution: when the sample becomes too small,
superconductivity breaks down.  More surprising was the finding
\cite{Golubev-94,vonDelft-96} that the breakdown is {\em parity-dependent}:
the odd $\Delta_1$ vanishes already at a much smaller level spacing (say
$d_1^{\rm GC}$) than the even $\Delta_0$ (say $d_0^{\rm GC}$).  This is
reflected in the condensation energy $E_p={\cal E}_{p}-\langle
F_p|H|F_p\rangle$, which is measured relative to the energy of the
respective uncorrelated Fermi sea ($|F_0 \rangle = \prod_{j<n_0}
c^\dagger_{j+} c^\dagger_{j-} |\mbox{Vac}\rangle$ or $|F_1\rangle =
c^\dagger_{n_0+} |F_0 \rangle$): Fig.~\ref{fig:results}(a) below shows that
the critical level spacing above which the g.c. results for $E_p^{\rm GC}$
reduce to zero is {\em parity-dependent.}\/

\subsection{Fluctuations}
While the g.c. BCS ansatz confirms Anderson's prediciton, it is,
however, clear that the approach has two problems: (a) it inherently
contains fluctuations of the particle number $N$ which are not present
in the actual grain system due to its large charging energy; and (b)
it fails to describe superconducting fluctuations for large level
spacings $d\gg\tilde\Delta$, where it trivially yields $\Delta=0$.

For bulk systems, the fluctuations in $N$ are negligibly small
($\delta N \sim \sqrt N \ll N$), and a more rigorous fixed-$N$
treatment would only correct the BCS ground state energy per electron
\cite{Anderson-58,Muehlschlegel-62} by order $1/N$, which vanishes in
the thermodynamic limit $d\to0$. For ultrasmall systems, however, 
precisely such corrections are important and we will have to
incorporate them.

On the other hand, pairing correlations (the redistribution of pairs from
below to above $\varepsilon_F$) can lower the condensation energy below zero
even when $\Delta_p = 0$.  In this regime, pairing correlations are
traditionally called ``superconducting fluctuations'', which are evidently not
captured adequately by the g.c. ansatz (\ref{eq:BCSground}). Matveev and
Larkin (ML) \cite{Matveev-97} calculated the energy lowering to logarithmic
order for $d\gg\tilde\Delta$ (i.e.\ $\Delta_{p}=0$). However, they are known
\cite{Golubev-94} to become important already in the crossover regime
$d\sim\tilde\Delta$.

To adequately describe an ultrasmall superconducting grain we
therefore must go beyond standard BCS theory by (a) fixing the
particle number $N$ and (b) incorporating superconducting
fluctuations. These ingredients will allow us to describe the full
crossover between the bulk system, which is dominated by BCS
superconductivity, and the few-electron system, which mainly shows
superconducting fluctuations. In particular we shall study how, at the
crossing of the two energy scales $d$ and $\tilde\Delta$, the
break-down of superconductivity predicted by BCS theory is softened
and how the fluctuations become dominant.

\section{The Full Crossover from Small $N$ to the Bulk Limit}
\subsection{Fixed-$N$ Projection}
Our first crossover study \cite{Braun-98b} adapts a method developed by
Dietrich, Mang and Pradal \cite{Dietrich-64} for shell modells of nuclei
with pairing interaction to the case of ultrasmall metallic grains. This
``projected BCS'' (PBCS) method also is a variational approach, but
projects (before variation) the trial wave-function onto a fixed electron
number $N=2n_0+p$ ($p$ is the parity):
\begin{eqnarray}
  \label{eq:dmp-wavefunction}
  \lefteqn{|\text{PBCS}\rangle_0 =} \nonumber \\
  & &{C \int_0^{2\pi}\!\!\!\! d\phi\,e^{-i\phi n_0} \prod_{j=0}^{N-1} 
    \Big(u_j+
     e^{i\phi} v_j c^\dagger_{j+}c^\dagger_{j-}\Big) |\mbox{Vac}\rangle,}
\end{eqnarray}
(In the odd case $\prod_j$ again is replaced by
$c^\dagger_{j_{\text{odd}}}\prod_{j\neq j_{\text{odd}}}$.)  The integral
over $\phi$ performs the projection onto the fixed electron pair number
$n_0$, and $C$ is a normalization constant ensuring
${}_0\langle\text{PBCS}|\text{PBCS}\rangle_0=1$. Again, the amplitudes
$v_j$ and $u_j$ are found by minimizing the energy expectation value of the
{\em projected\/} wave-function
\begin{math}
  {\cal E}^{\text{PBCS}}_p = {}_p\langle\text{PBCS}|\hat H|\text{PBCS}\rangle_p.
\end{math}
While in the g.c. case the wave-function essentially can be described by a
single parameter $\Delta_{p}$, in this case the minimization leads to a
{\em set\/} of $2n_0$ coupled non-linear equations which include projection
integrals (for details see \cite{Braun-98b}).  Following
Ref.~\cite{Dietrich-64}, we evaluate all integrals numerically (using fast
Fourier transform routines).

In the limit $d\to0$ at fixed $n_0d$, the PBCS theory reduces to the g.c.\ 
BCS theory of Ref.~\cite{Braun-97} (proving that the latter's
$N$-fluctuations become negligible in this limit): The projection integrals
can then be approximated by their saddle point values \cite{Dietrich-64};
at the saddle, the variational equations decouple and reduce to the BCS gap
equation while the saddle point condition fixes the {\em mean\/} number of
electrons to be $2n_0$. To check the opposite limit of $d\gg\tilde\Delta$
where $n_0$ becomes small, the so-called fluctuation-dominated regime, we
compared the PBCS results for $E_0$ with exact results, finding agreement
to within, say,
$6\%$ for $n_0 \le 10$.%
\footnote{Because this error refers to the {\em correlation energy},
  it is larger than the error ($< 1\%$) cited in \cite{Braun-98b} for
  the total ground state energy ${\cal E}_p$.}  This shows that
superconducting fluctuations are automatically treated adequately in
the PBCS approach.

The advantages of the PBCS method relative to the g.c. one are (a) the
similarity of the trial wave-function to the BCS wave function,
allowing it to 
capture the bulk limit and, on the other hand, (b) the increase of variational
degrees of freedom from one ($\Delta$ in the g.c. BCS theory) to $N-1$ (one
for each single particle level minus 1 for fixing $N$) due to the projection
onto a fixed particle number. These additional degrees of freedom allow the
method to also capture the superconducting fluctuations for
$d\gg\tilde\Delta$.  Because it works so well for $d \ll \tilde\Delta$ and
$d\gg\tilde\Delta$, one might hope that it acceptably describes the crossover
regime $d\sim\tilde\Delta$, too. There it really amounts to an
uncontrolled approximation, whose quality will be checked against
exact results in the next section. 

\subsection{Richardson's Exact Solution}
Since the publication of \cite{Braun-98b} we became aware of the fact that
Richardson had shown already in the mid-60's \cite{Richardson}, also in the
context of nuclear physics, that the ground state of the Hamiltonian
(\ref{eq:hamiltonian}) actually can (for non-degenerate $\epsilon_j$) be
found {\em exactly\/} by solving a set of $n_0$ coupled algebraic
non-linear equations.\footnote{The numerical implementation of the exact
  solution is much easier than for the PBCS method, since it does not
  include any projections integrals.}

Richardson introduces electron pairs $b_j = c_{j-}c_{j+}$, which have
the commutation relation $[b_j, b_{j'}^\dagger]=\delta_{jj'}(1-2b_j^\dagger
b_j)$, and exploits the fact that in the Hilbert space of 
non-singly-occupied states, the modified Hamiltonian
\begin{eqnarray}
  \label{eq:rich-hamiltonian}
  \tilde H = \sum_j 2\epsilon_j b_j^\dagger b_j - \lambda d \sum_{jj'}
  b_j^\dagger b_{j'} \; . 
\end{eqnarray} 
is equivalent to the $H$ of (\ref{eq:hamiltonian}).  Solving $\tilde H$
exactly would be trivial if the $b$'s would represent true bosons.
However, they actually are ``hard-core bosons'' instead. Richardson thus
expresses the general ground state $|G\rangle^N$ of the system as
\begin{eqnarray}
  |G\rangle^N = \sum_{j_1\neq\cdots\neq j_{n_0}} \varphi(j_1\cdots
  j_{n_0}) b^\dagger_{j_1}\cdots b^\dagger_{j_{n_0}}|\text{Vac}\rangle,
\end{eqnarray}
where the sum is explictly restricted to exclude double occupancy of pair
states.  The wave-function $\varphi$ is found by solving the many-body
Schr\"o\-di\-nger equation for $\varphi$. Richardson showed that the
following ansatz works:
\begin{eqnarray}
  \label{eq:ansatz}
  \varphi(j_1\cdots j_{n_0}) \propto \sum_{{\cal P}} {\cal P}\left\{\prod_{k=1}^{n_0} \frac1{2\epsilon_{j_k}-E_{{\cal P}(k)}}\right\}.
\end{eqnarray}
Here $\sum_{\cal P}{\cal P}$ represents the sum over all permutations of
$1,\ldots, {n_0}$, and the parameters $E_k$ are the solution of the coupled
algebraic equations
\begin{eqnarray}
  \label{eq:richardson}
  \frac1{\lambda d}+\sum_{{l=1 \atop l\neq k}}^{n_0}\frac2{E_l-E_k} & = & 
  \sum_{j=1}^{2n_0} \frac1{2\epsilon_j-E_k},\qquad k = 1\ldots n_0.
\end{eqnarray}
The total ground state energy is given by
\begin{math}
  {\cal E}^{\text{exact}}_0 = \sum_{i=1}^{n_0} E_i.
\end{math}

The discussion of Richardson's exact solution in context with ultrasmall
grains will be the subject of a forthcoming paper.

\subsection{Ground State Energies}
\begin{figure}[p]
  \begin{center}
    \leavevmode
    \psfig{figure=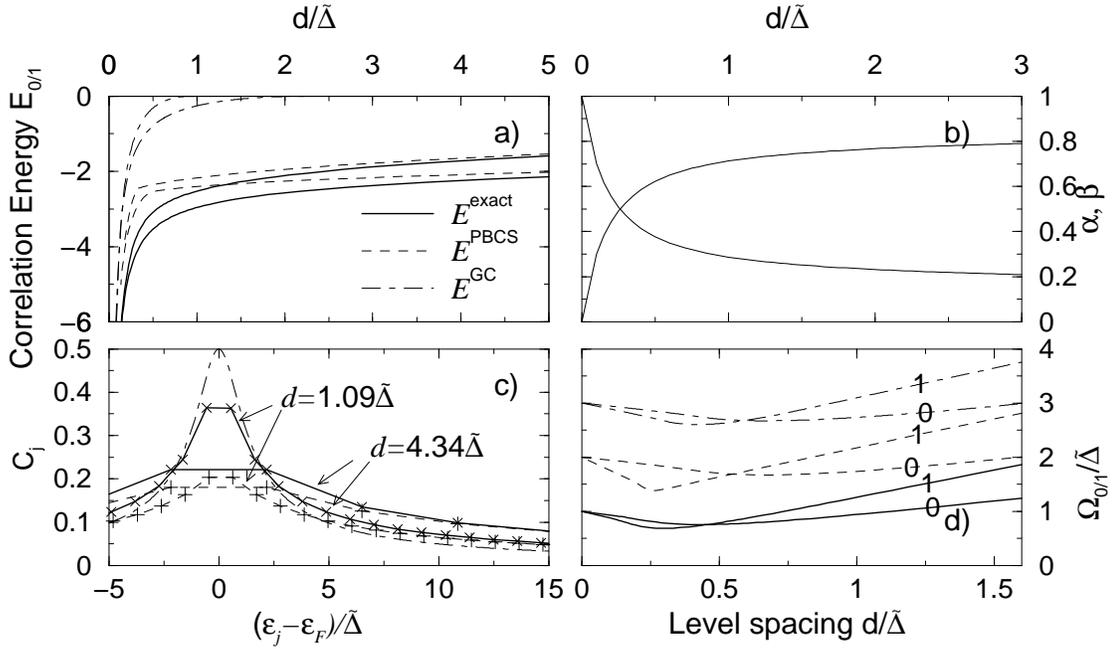,width=0.8\hsize}
    \caption{(a) The ground state correlation energies ${E}_p$ for even
      and odd systems (even: lower curves, odd: upper curves), calculated
      grand-canonically (GC), with PBCS and Richardson's method (exact) as
      functions of $d/ \tilde \Delta = 2 \sinh (1/ \lambda) / N$.  (b) shows
      the coefficients $\alpha$ and $\beta$ of Eq. (\ref{eq:fit}) as discussed
      in the text. The pairing amplitudes $C_j$ of Eq.~(\protect\ref{eq:Cj})
      are shown in (c) for bulk (no symbols, dot-dashed), 
  PBCS (``$+$'', dashed) and exact results
      (``$\times$'', solid) for  $d=1.09\tilde\Delta$ and
      $d=4.34\tilde\Delta$. For $d=0.27\tilde\Delta$,
 the PBCS and exact curves (not shown) are indistinguishable from
the bulk curve.
(d), finally, shows the results the pair-breaking
      energies $\Omega_{0/1}$ as calculated with g.c. BCS (dot-dashed) PBCS
      (dashed) and exactly (solid). (For clarity the graphs are
      offset by successively one unit each.)}
    \label{fig:results}
  \end{center}
\end{figure}

Figure~\ref{fig:results}(a) shows the ground state condensation energies
for both even and odd grains calculated with g.c. BCS method, the PBCS
approach (for $N\le600$) and Richardson's exact solution. The result
$E_{b}^{\text{GC}}$ \cite{Braun-99} is also shown for comparison.  The
g.c.\ curves suggest the aforementioned ``breakdown of superconductivity''
\cite{vonDelft-96,Braun-97} at some critical $p$-dependent level spacing
$d^{\text{GC}}_{p}$ above which $E_{p}^{\text{GC}}=0$.  In contrast, the
${E}_p^{\text{PBCS}}$'s (i) are significantly lower than the
$E^{\text{GC}}_p$'s, thus the projection much improves the variational
ansatz; and (ii) are negative for {\em all}\/ $d$, which shows that the
system can {\em always} gain energy by allowing pairing correlations, even
for arbitrarily large $d$. The exact solution $E^{\text{exact}}_p$ further
improves the PBCS results, especially for intermediate level spacings.  The
PBCS results are evidently quite accurate for $d\gg\tilde\Delta$ and, like
the g.c.  results, for $d\ll\tilde\Delta$.  

As anticipated in \cite{Braun-99}, the ``breakdown of superconductivity''
is evidently not as complete in the canonical as in the g.c.\ case.
Nevertheless, some remnant of it does survive in $E_{b}^{\text{PBCS}}$,
since its behaviour, too, changes markedly at a $p$ (and $\lambda$)
dependent characteristic level spacing $d_{p}^C$ ($< d^{\text{GC}}_p$): it
marks the end of bulk BCS-like behavior for $d<d_p^C$, where ${ E}_p^{C}$
is {\em extensive}\/ ($\sim 1/d$), and the start of a
fluctuation-dominated
\ plateau for
$d>d_p^C$, where $E_{p}^{C}$ is practically {\em intensive\/} (almost $d$
independent). 

The standard heuristic interpretation \cite{Tinkham-2} of the
bulk BCS limit $-\tilde\Delta^2/(2d)$ (which is indeed reached by $E_p^C$
for $d\to 0$) hinges on the scale $\tilde \Delta$: the number of levels
strongly affected by pairing is roughly $\tilde \Delta / d$ (those within
$\tilde \Delta$ of $\epsilon_F$), with an average energy gain per level
of $-\tilde\Delta/2$. To analogously interpret the $d$ {\em
  in}\/dependence of $E_p^{C}$ in the fluctuation-dominated 
regime, we argue that {\em the
  scale $\tilde \Delta$ loses its significance}\/ -- fluctuations affect
{\em all}\/ $n_0 = \omega_D/d$ unblocked levels within $\omega_D$ of
$\epsilon_F$ (this is made more precise below), and the energy gain per
level is proportional to a renormalized coupling $- \tilde \lambda d$. 

The exact results smear out the crossover even more than the PBCS results
($E_p^{\rm exact}$ lacks the kinks of $E_p^{\rm PBCS}$), so much so that no
sharply-defined crossover level spacing can be associated with $E_p^{\rm
  exact}$.  However, the crossover scale evidently still is $d \sim \Delta$.
This can be confirmed by analyzing the functional dependence of the ground
state energy on the coupling strength $\lambda$: In the BCS limit, $E_p\approx
-\tilde\Delta^2/(2d)$, where $\tilde\Delta$ depends exponentially on $\lambda$
[since $\tilde \Delta = \omega_D \sinh(1/\lambda)$].  In the
fluctuation-dominated regime, however, perturbation theory in $\lambda$
suffices and the correlation energy is roughly linear in $\lambda$.  For each
$d$, we thus fitted the numerical results for $E^{\text{exact}}_p(d,\lambda)$,
calculated for various $\lambda$, to
\begin{eqnarray}
  \label{eq:fit}
  {E_0(\lambda,d) \over E_0(\lambda_0,d)} =  
\alpha(d)\frac{\sinh(1/\lambda_0)^2}{\sinh(1/\lambda)^2}
    - \beta(d) \frac{\lambda}{\lambda_0} ,
\end{eqnarray}
a ``phenomenological ansatz'' which intends to capture the relative importance
of the exponential or linear $\lambda$ dependence in the coefficients
$\alpha(d)$ and $\beta(d)$.  The results, shown in Fig.~\ref{fig:results}(b),
clearly show the crossover from the BCS-dominated regime ($\alpha>\beta$) to
the fluctuation-dominated regime ($\alpha<\beta$).

\subsection{Wave Functions}
Next we analyze the ground state wave-function, which 
can be characterized by 
\begin{equation}
  \label{eq:Cj}
  C^2_j (d) =\langle c^\dagger_{j+} c_{j+} c^\dagger_{j-} c_{j-}\rangle
  -\langle c^\dagger_{j+} 
  c_{j+} \rangle\langle c^\dagger_{j-} c_{j-}\rangle, 
  \label{C_j}
\end{equation}
a set of correlators that measure the amplitude enhancement for finding a
{\em pair\/} instead of two uncorrelated electrons in a single-particle
niveau $|j,\pm\rangle$.  For all $j$ of an uncorrelated state one has
$C_j=0$.  For the g.c.\ BCS case $C_j = u_j v_j$ and the $C_j$'s have a
characteristic peak of width $\sim \tilde\Delta$ around $\epsilon_F$, see
Fig.~\ref{fig:results}(c), implying that pairing correlations are
``localized in energy space''.  For the BCS regime, both canonical methods
produce $C_j$'s virtually identical to the g.c.\ case, {\em vividly
  illustrating why the g.c.\ BCS approximation is so successful: not
  performing the canonical projection hardly affects the parameters $v_j$
  if $d \ll \tilde\Delta$, but tremendously simplifies their calculation}.
However, in the fluctuation-dominated regime
$d\,{\stackrel{\scriptscriptstyle >}{\scriptscriptstyle
    \sim}}\,\tilde\Delta$, the character of the wave-function changes:
weight is shifted into the tails far from $\epsilon_F$ at the expense of
the vicinity of the Fermi energy.  Thus {\em pairing correlations become
  delocalized in energy space} (as also found in \cite{Mastellone-98}), so
that referring to them as mere ``fluctuations'' is quite appropriate. In
the extreme case $d \gg \tilde\Delta$ superconducting fluctuations are
roughly equally strong for all interacting levels.

\subsection{Parity Effect}

The parity effect predicted in the g.c. ensemble can be studied with a simple
generalization \cite{Braun-99} of the above methods. Specifically, we shall
study the parity-dependent ``pair-breaking energy'' $\Omega_p$, i.e. the
minimum energy required to break a pair by flipping a spin in an even or odd
grain, defined as $\Omega_0=\frac12(E_2-E_0)$ and $\Omega_1=\frac12(E_3-E_1)$,
where $E_u$ denotes the energy of the lowest-lying state with $u$ unpaired
electrons with the same spin. (The pair-breaking energies can readily be
measured in BRT's experiments by applying a magnetic field, whose Zeeman
energy favors the breaking of pairs.) For a non-interacting system and
approximatly also for large $d\gg\tilde\Delta$, $\Omega_0\simeq d/2$ while
$\Omega_1\simeq d$. On the other hand, in the bulk limit $\Omega_0 = \Omega_1=
\tilde \Delta$, since in the bulk $E_{p+2}=E_p+2\tilde\Delta$. The parity
effect now states that as $d$ increases from $d \simeq 0$, the pairing
correlations die faster for an odd than an even grain, causing $\Omega_1$ to
initially decrease faster than $\Omega_0$.  Since in the large $d$-limit
$\Omega_1>\Omega_0$, the two energies must cross somewhere at
$d\sim\tilde\Delta$, as shown in Fig.~\ref{fig:results}(d). This crossing is a
manifestation of the parity effect already predicted in the g.c. framework.

Remarkably, despite the crudeness and incorrect treatment of
fluctuations of the g.c. method, it gives surprisingly good results
for energy differences like the $\Omega_p$.  The fluctuations which it
neglects seem to cancel substantially in energy differences such as
$\Omega_p$.  This observation {\it a posteriori\/} justifies the use
of the g.c. method even for small grains, at least for rough
calculations of energy differences.

\section{Conclusion}

In summary, the crossover from the bulk to the fluctuation-dominated
 regime can be captured in
full using a fixed-$N$ projected BCS ansatz, or even exactly using Richardson's
method.  With increasing $d$, the pairing correlations change from being
strong and localized within $\tilde\Delta$ of $\epsilon_F$, to being mere
weak, energetically delocalized ``fluctuations''; this causes the condensation
energy to change from being {\em extensive\/} to {\em intensive\/} (modulo
small corrections). Thus, the qualitative difference between
``superconductivity'' for $d\,{\stackrel{\scriptscriptstyle
    <}{\scriptscriptstyle \sim}}\,\tilde\Delta$, and ``fluctuations'' for $d
\,{\stackrel{\scriptscriptstyle >}{\scriptscriptstyle \sim}}\, \tilde\Delta$,
is that for the former but not the latter, adding {\em more\/} particles gives
a {\em different\/} condensation energy; for superconductivity, as Anderson
put it, ``more is different''.

\widetext
\vspace*{5ex}
\noindent {\em Note added:\/}\\
After submission of this manuscript works by Dukelsky and Sierra
\cite{Dukelsky-99,Dukelsky-99b} used DMRG methods to numerically calculate
the ground states of the presented Hamiltonian (\ref{eq:hamiltonian}).
Their results agree within a relative error of $10^{-4}$ with those
obtained by Richardson's exact method.

\end{document}